\documentstyle[aps,epsf]{revtex}
\begin{document}
\title{Statistical Models of Nuclear Fragmentation}
\author{Scott Pratt and Subal Das Gupta\footnote{Permanent address: Physics
Department, McGill University, Montreal, Quebec, Canada, H3A 2T8 }}
\address{Department of Physics and Astronomy and\\
National Superconducting
Cyclotron Laboratory,\\
Michigan State University, East Lansing, MI 48824~~USA}
\date{\today}
\maketitle

{\abstract A method is presented that allows exact calculations of fragment
multiplicity distributions for a canonical ensemble of non-interacting
clusters. Fragmentation properties are shown to depend on only a few
parameters. Fragments are shown to be copiously produced above the transition
temperature. At this transition temperature, the calculated multiplicity
distributions broaden and become strongly super-Poissonian. This behavior is
compared to predictions from a percolation model. A corresponding
microcanonical formalism is also presented.}


\section{Introduction}

Heavy ion collisions where excitation energies are of the order of 10 MeV per
nucleon probe the energy regime where the nuclear liquid gas transition is
expected to take place. Below energies of approximately 50$A$ MeV, symmetric
collisions are expected to produce sources that evaporate particles as would be
expected from a hot liquid drop, whereas above this threshold, the excited
source is expected to explode, producing larger clusters through simultaneous
multi-fragmentation. In this energy regime, the process of fragment production
is not clear, and comparisons with data have been made with a disparate set of
models, ranging in simplicity from percolation descriptions\cite{bauer,campi}
and lattice gas models\cite{pan1}, to evaporative models\cite{friedman},
dynamical simulations\cite{chomaz,ohnishi,feldmeier,kiderlen,montoya}, and
microcanonical samplings\cite{randrup,copenhagen}.

Fluctuations behave in a special manner in the region of phase transition, so
it should seem that the study of fluctuations of fragmentation observables
might prove insightful for investigating multifragmentation. Moretto and
collaborators\cite{moretto} have measured multiplicity distributions of
intermediate-mass fragments (IMFs) as a function of excitation energy for a
variety of projectile/target combinations utilizing beams with energies up to
60 $A$ MeV. The analysis showed sub-Poissonian multiplicity distributions that
could accurately be described with binomial distributions. These observations
have inspired a variety of explanations\cite{toke,tsang_criticism,bauer_size,morettoreply}.

Recently, Chase and Mekjian\cite{chase} have discovered a method for exact
calculation of the canonical partition function for non-interacting
clusters. In this paper we extend this approach to include fragmentation
observables. We present a method for exact determination of both multiplicity
distributions and their moments. When raising the temperature while keeping the
volume fixed, we observe a sharp transition for fragmentation at the same point
where the specific heat peaks. At this threshold, the multiplicity distribution
becomes remarkably wide. We associate this behavior with a first order phase
transition and remark that the broadening might well disappear in a
microcanonical treatment.

We parameterize the width of the fragment multiplicity distribution, relative
to the mean, with a correlation coefficient $\xi$ which is described in the
next section. The method for calculating fragmentation observables from a
canonical ensemble is presented in section \ref{methods_sec} while the results
are presented in section \ref{results_sec}. In section \ref{perc_sec} we
contrast the results of this model with those of a percolation
model. Expressions that could be used for microcanonical calculations are
presented in section \ref{microcanonical_sec}.

\section{Multiplicity distributions and the correlation coefficient}
\label{generalproperties_sec}

A super or sub-Poissonian multiplicity distribution is one whose variance
exceeds or falls below it's mean respectively. The difference of the variance
and the mean can also be written as a correlation. We demonstrate this by
considering the emission into an arbitrarily large number of states $i$, each
of which is infinitesimally probable. A state could be defined as a specific
type of IMF emitted into an arbitrarily small bin in momentum space. The
difference of the variance and the mean is
\begin{eqnarray}
\sigma^2-\langle n\rangle&=&\sum_{i,j}(\langle n_i-\langle n_i\rangle)
(n_j-\langle n_j\rangle)-\sum_i\langle n_i\rangle\\
\nonumber
&=&\sum_{i\ne j}(\langle n_i-\langle n_i\rangle)(n_j-\langle n_j\rangle)
+\sum_i (\langle n_i-\langle n_i\rangle)^2-\langle n_i\rangle
\end{eqnarray}
If the bins are arbitrarily small, one may discard the terms in the second sum
proportional to $<n_i>^2$, then use the fact that $n_i^2=n_i$ for $n_i$=0
or 1, to discard the remainder of the second term and obtain,
\begin{equation}
\label{xidef_eq}
\xi\equiv \frac{\sigma^2-\langle n\rangle}{\langle n\rangle^2}
=\frac{\sum_{i\ne j}(\langle n_i-\langle n_i\rangle)
(n_j-\langle n_j\rangle)}{\langle n\rangle^2}
\end{equation}
Here, $\xi$ can be interpreted as a correlation coefficient. It is positive if
the emission into two different bins is positively correlated. The only
assumption going into this derivation is that the bins may be divided into
arbitrarily small sizes.

The simplest examples of sub- and super-Poissonian multiplicity distributions
are the binomial and negative binomial distributions. The binomial distribution
is defined in terms of two parameters, $p$ and $N$.
\begin{equation}
P_n=\frac{N!}{n!(N-n)!}p^n(1-p)^{N-n},
\end{equation}
where $p$ is the probability of success in one of $N$ tries. For the binomial
distribution the mean and correlation coefficients become
\begin{eqnarray}
\langle n\rangle_{\rm bin.}&=&pN\\
\xi_{\rm bin.}&=&-1/N,
\end{eqnarray}
and stays negative. Like most correlations it is proportion to the inverse of
the system size.

The negative binomial distribution is also defined by two parameters $p$ and
$N$.
\begin{equation}
P_n=\frac{(N+n-1)!}{(N-1)!n!}\frac{p^n}{(1+p)^{N+n}}.
\end{equation}
The correlation coefficient in this case is opposite to that of the binomial
distribution.
\begin{eqnarray}
\langle n\rangle_{\rm neg. bin.}&=&pN\\
\xi_{\rm neg. bin.}&=&1/N.
\end{eqnarray}
Binomial and negative binomial distributions result when one considers
populating $N$ quantum levels with fermions or bosons respectively with $p$
representing the average population of each level.

Random emission from a large number of uncorrelated sources leads to a
Poissonian distribution. The binomial distribution suggests that conservation
of particle number would give a negative correlation coefficient of order
$1/N$, where $N=A/a$ is the number of IMFs of characteristic size $a$ that
could fit into the system. For sufficiently small systems, this negative
contribution from particle-number conservation dominates, with the extreme case
being where $a$ is more than half the system size meaning that no more than one
IMF can be emitted. Other negative correlations are expected due to energy
conservation. If IMF emission requires energy, e.g. escaping a Coulomb barrier,
energy conservation is expected to reduce the probability of emitting a second
IMF. We will see in the next sections that positive correlations can arise,
principally due to surface considerations.

Aside from the size of the correlations, it is also of interest to understand
whether the entire multiplicity distribution, is well described by a
two-parameter fit to a binomial or negative binomial distribution. If the
reduction of the results to two parameters is valid, comparison of different
models and data is greatly simplified.

\section{Canonical distributions of non-interacting clusters}
\label{methods_sec}
Chase and Mekjian have shown that canonical partition functions can be easily
calculated in terms of the partition functions of single clusters. This allows
the calculation of thermodynamic quantities for a system of fixed nucleon
number without resorting to numerically intensive Monte-Carlo procedures. If
the partition function for a single cluster of size $a_k$ is denoted by
$\omega_k$, the partition function for a system of size $A$ may be written,
\begin{eqnarray}
\label{recursion_eq}
\Omega_A&\equiv&\sum_{\langle\Sigma n_ka_k=A\rangle}
\prod_k \frac{\omega_k^{n_k}}{n_k!}\\
\nonumber
&=&\sum_k \omega_k\frac{k}{A}\Omega_{A-a_k}
\end{eqnarray}
Thus, $\Omega_A$ is expressed recursively in terms of $\omega_k$ and
$\Omega_{A^\prime}$ for $A^\prime<A$. Proof of this relation is given in the
appendix. The only shortcoming of this approach is that explicit (not mean
field) interactions between fragments are ignored.

Moments of the multiplicity distribution may be expressed in terms of the
partition functions. The moments can then be used to derive the correlation
coefficient, $\xi_a$ defined in Eq. (\ref{xidef_eq}) or the multiplicity
distribution as discussed below. The first moment is the mean which is defined
as:
\begin{equation}
\langle n_k\rangle=\omega_k\frac{\Omega_{A-a_k}}{\Omega_A}
\end{equation}

Rather than considering moments, $\langle n_b^m\rangle$, it is more convenient
to consider factorial moments, $F_{b,A,m}$, defined as:
\begin{equation}
\label{factmomdef_eq}
F_{b,A,m}\equiv\langle n_b(n_b-1)\cdots(n_b-m+1)\rangle\\
\end{equation}
Calculation of the factorial moments for $n_b$ defined in
Eq. (\ref{factmomdef_eq}}) is simple if $b$ refers to single species $k$:
\begin{equation}
\label{singlespeciesfactmom_eq}
F_{k,A,m}=\omega_k^m\frac{\Omega_{A-ma_k}}{\Omega_A}.
\end{equation}
However, if $b$ refers to a set of species, $n_b=\Sigma_{k\in b} n_k$, where
the various species that comprise $b$ have different masses,
Eq. (\ref{singlespeciesfactmom_eq}) is no longer valid. One must then generate
the factorial moments using the recursion relation,
\begin{equation}
\label{factmom_recursion_eq}
F_{b,A,m}=\sum_{k\in b} \omega_kF_{b,A-a_k,m-1}
\frac{\Omega_{A-a_k}}{\Omega_A},
\end{equation}
which is true in general. The proofs of Eq. (\ref{singlespeciesfactmom_eq}) and
Eq. (\ref{factmom_recursion_eq}) are given in the second section of the
appendix.

As shown in the third subsection of the appendix, factorial moments are
sufficient for calculating the entire multiplicity distribution via the
relation,
\begin{equation}
\label{multdistfromfactmom_eq}
P_{b,A,n}=\sum_{m\ge n}F_{b,A,m} \frac{1}{(m-n)!n!}(-1)^{m-n},
\end{equation}
where $P_{b,A,n}$ gives the probability of viewing $n$ fragments of type $b$ in
a system of size $A$. 

More directly, one may also generate the multiplicity distribution, without
knowing the factorial moments, through the recursion relation,
\begin{equation}
\label{multdist_recursion_eq}
P_{b,A,n}=\frac{1}{n}\sum_{k\in b}\omega_k 
P_{b,A-a_k,n-1}\frac{\Omega_{A-a_k}}{\Omega_A}.
\end{equation}
Proof of Eq. (\ref{multdist_recursion_eq}) is presented in the appendix. 
This direct method of producing the multiplicity distribution has proven more
numerically stable than generating the distribution from the factorial moments.
This improvement can be traced to the alternating signs in
Eq. (\ref{multdistfromfactmom_eq}).

Summarizing the technique, one starts by calculating partition functions for
individual fragments $\omega_k$. One may then generate partition functions,
$\Omega_A$, by using the recursion relation, Eq. (\ref{recursion_eq}). The
recursion relation for factorial moments, Eq. (\ref{factmom_recursion_eq}),
then allows one to generate the factorial moments, which in turn allow the
determination of the entire multiplicity distribution using
Eq. (\ref{multdistfromfactmom_eq}). Alternatively, one may calculate the
multiplicity distributions directly using
Eq. (\ref{multdist_recursion_eq}). The obtained multiplicity distributions are
exact. Although the sums used in the recursion relation are performed
numerically, they require only a fraction of a second of computer time.

\section{A liquid-drop picture for individual clusters}
\label{results_sec}

For our purposes, we consider the partition function for individual fragments
of mass $k$ as,
\begin{eqnarray}
\omega_k&=&V \left\{\frac{a_kMT}{2\pi}\right\}^{3/2}
e^{-F_{k,int}/T},\\
F_{k,int}&=&f_ba_k+f_sa_k^{2/3}+f_c\frac{1}{4}k^{5/3},
\end{eqnarray}
where the volume of the system is $V$, the mass of a single nucleon is $M$, and
the fragment's internal free energy $F_{k,int}$ is split into a bulk term and a
surface term. One sees that the bulk term is irrelevant in determining
fragmentation observables since it factors out of the partition function. Thus,
aside from the system size $A$, all fragmentation observables are determined by
three parameters, the ratio of the surface term to the temperature $f_s/T$, the
intrinsic entropy, $s\equiv(V/A)(MT)^{3/2}$, and Coulomb term, $f_c$. This
implies that many details of a system's microscopic makeup, e.g. Fermi vs. Bose
nature of the internal excitation, are irrelevant in determining the statistics
of fragmentation. If the surface term is negligible fragmentation is determined
purely by the intrinsic entropy.

For the surface and Coulomb terms, we use the parameters of the nuclear
liquid-drop model,
\begin{eqnarray}
f_s&=&17.2 {\rm MeV}\\
f_c&=&0.70\left(1-\left(\frac{\rho}{\rho_0}\right)^{1/3}\right){\rm MeV},
\end{eqnarray}
where the form of the Coulomb term was taken to account for the screening of
the Coulomb repulsion by the nuclear medium in a Wigner-Seitz like
parameterization\cite{copenhagen} with $\rho_0$ referring to nuclear saturation
density of 0.15 fm$^{-3}$.

All the calculations presented in this paper assumed a density of one third of
nuclear matter density. The behavior at different densities is not
qualitatively different, with the exception of the relative importance of the
Coulomb term. The fragmentation transition described below occurs when
$s=(1/\rho)(mT)^{3/2}$ is of order unity. Therefore a change of density affects
the temperature where fragmentation sets in. An excluded volume could easily be
incorporated into the problem by replacing $V$ with $V(1-\rho/\rho_0)$. This is
equivalent to changing the density, and does not qualitatively affect the
results. The surface energy is chosen to be a constant, $f_s=17.2$ MeV. One
could imagine scaling $f_s$ as a function of $\rho$ or temperature, although
one might object to incorporating a temperature dependence that is not of the
nature, $e^{E/T}$. The choice of $f_s$ does affect the transition temperature
and it's width. Larger choices of $f_s$ lead to sharper transitions. The
surface term should disappear at the critical point.  Although it is easy
to incorporate excluded volume effects and a running surface term, we choose
not to, as our principal goal in this study is to understand the general
properties of this approach.

\begin{figure}
\epsfxsize=0.75\textwidth \centerline{\epsfbox{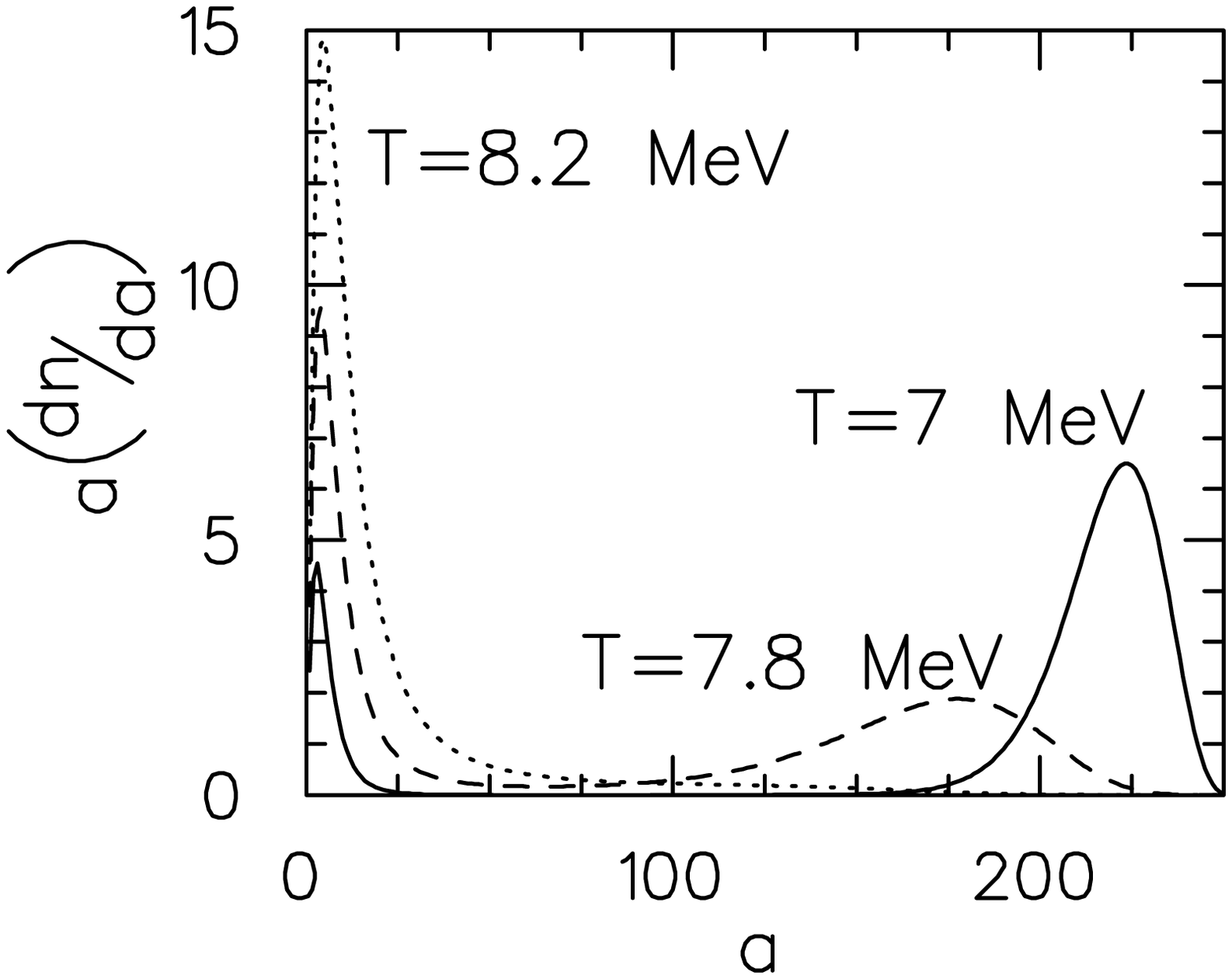}}
\caption{Mass distributions are displayed (scaled by $a$) at three temperatures
for an $A=250$ system. At the high temperature, 8.2 MeV, (dotted line) the
distribution is dominated by small fragments, while below the fragmentation
threshold, at 7 MeV, (solid line), most nucleons reside in one fragment. At 7.8
MeV (dashed line), the system seems equally divided between the large cluster
and small clusters. The transition occurs over a remarkably narrow range of
temperatures.
\label{massdist_fig}}
\end{figure}

In Figure \ref{massdist_fig}, the mass distribution $dN/da$ is shown for three
temperatures, $T$=7.0, 7.8 and 8.6 MeV.  The overall size of the system was
chosen to be 250 and Coulomb effects were neglected in these calculations. The
mass distribution has been multiplied by $a$ to emphasize how the composite
nucleons are partitioned into the various sized drops. At 7 MeV, the nucleons
are largely contained in one drop, while at the higher temperature, nearly all
the particles are part of small clusters.

\begin{figure}
\epsfxsize=0.75\textwidth \centerline{\epsfbox{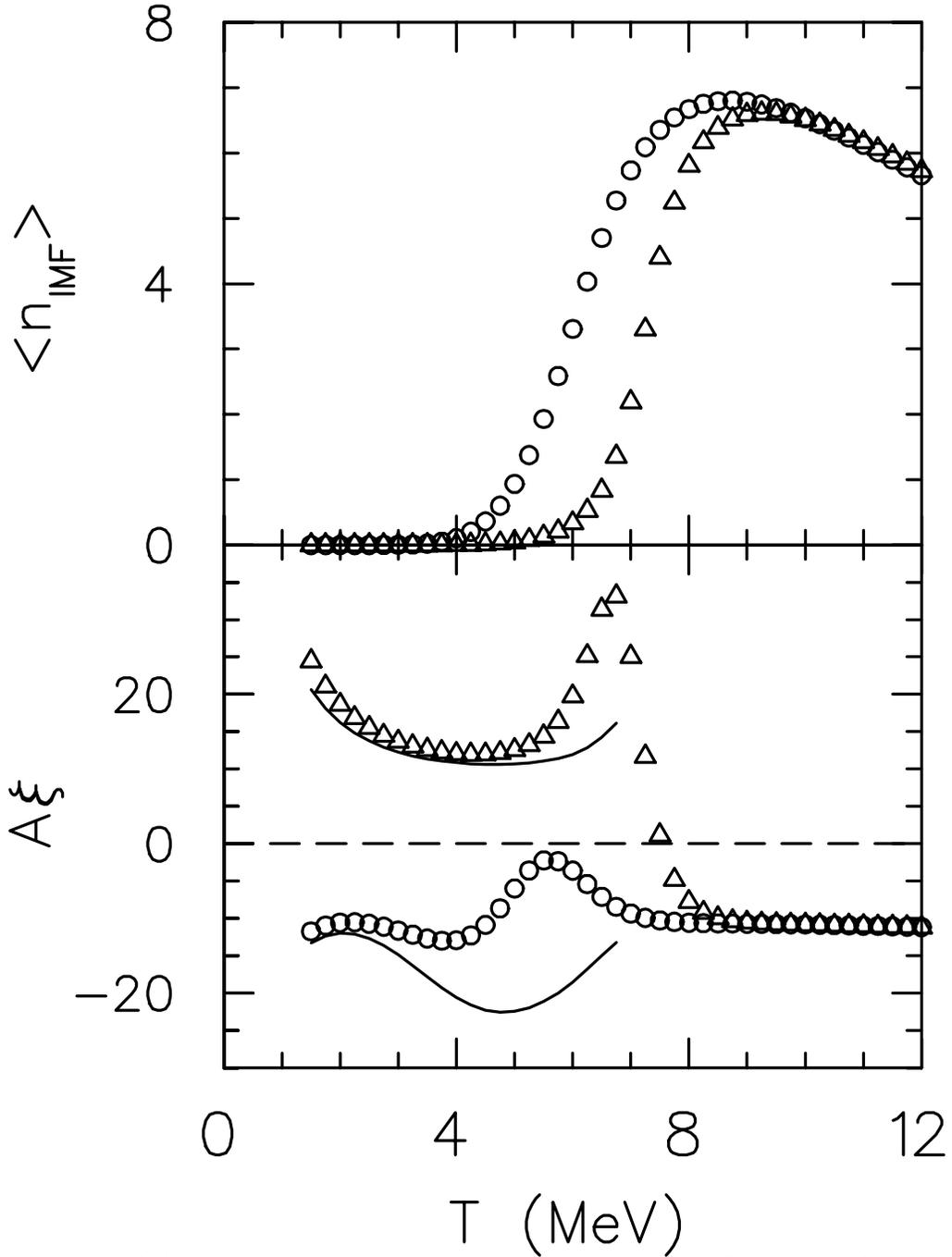}}
\caption{The average multiplicity of IMFs, those fragments with masses from 6
to 40, are shown in the upper panel as a function of temperature for an $A=100$
system. Including Coulomb (circles) lowers the threshold temperature for
fragmentation compared to when Coulomb is ignored (triangles). Correlation
coefficients are displayed in the lower panel. The approximations described in
the text, which are represented by lines, describe the behavior at low
excitation. A positive correlation at the fragmentation threshold arises from
two classes of events, those with and without a large cluster, both
contributing at the fragmentation threshold.
\label{meanxi_fig}}
\end{figure}

The average multiplicity of intermediate-mass fragments ($5< a\le 40$),
denoted as IMFs, are shown in the upper panel of Figure \ref{meanxi_fig}
for the cases where the density is 0.05 fm$^{-3}$ and the overall system size
is again $A=100$. Results are plotted against the temperature, both for the
case where the Coulomb term is included as well as for the case where it is
neglected. The inclusion of Coulomb pushes fragmentation down towards lower
temperatures. The trend would strengthen if we were to consider larger
systems. When the excitation energy exceeds the fragmentation threshold,
average IMF multiplicities quickly climb to over a half dozen per event.

Correlation coefficients are shown in the lower panel of Figure
\ref{meanxi_fig} with the system size is set at 100. Coefficients are
shown both for the case where Coulomb is included and for the case where
Coulomb is neglected.

When Coulomb is neglected super-Poissonian behavior ensues at lower excitations
as evidenced by the positive values of $\xi$. At low temperatures, this
behavior may be understood by considering the surface penalty for emitting a
single fragment of mass $a$,
\begin{equation}
\Gamma(A\rightarrow A-a)\propto 
\exp \left\{-a~\frac{d}{dA}(F_{\rm surf}/T)\right\}
\end{equation}
If a second IMF is emitted, it feels the same penalty, except for the fact that
$dF/dA$ is evaluated at a smaller overall size, $A-a$. The correlation thus
becomes,
\begin{eqnarray}
\xi_{\rm surf}&\approx&\frac{\Gamma(A\rightarrow A-a)\Gamma(A\rightarrow A-a)}
{\Gamma(A-a\rightarrow A-2a)^2}\\
\nonumber
&\approx&\frac{a^2}{T}\frac{d^2F_{\rm surf}}{dA^2}\\
\nonumber
&=&\frac{2a^2f_s}{9T}A^{-4/3}.
\end{eqnarray}
The dashed line represents the surface contribution to $\xi$ and closely
follows the statistical calculation at lower temperatures. The average size of
an IMF was used for $a$.

The inclusion of Coulomb reduced $\xi$ to negative values as seen in Figure
\ref{meanxi_fig}. The Coulomb contribution to $\xi$ may be approximated in a
similar fashion as the surface contribution,
\begin{equation}
\xi_{\rm coul}\approx-\frac{5a^2f_c}{9T}A^{-1/3}.
\end{equation}
Again, the approximation is plotted as a dashed line in the lower panel of
Figure \ref{meanxi_fig}.

The simple estimate of $\xi$ works well at low temperature, where most of the
particles reside in a single fragment. The approximations for $\xi$ improve for
larger systems when Coulomb is ignored but become worse for large systems when
Coulomb is included. This arises because a large system does not wish to form a
single drop when Coulomb is included.

At high temperatures the sub-Poissonian behavior may be roughly understood as
arising from particle conservation. As seen in section
\ref{generalproperties_sec} when considering binomial distributions, one
expects a negative correlation of order $1/N$, where $N$ is the maximum number
of IMFs that fit into the system.

\begin{figure}
\epsfxsize=0.75\textwidth \centerline{\epsfbox{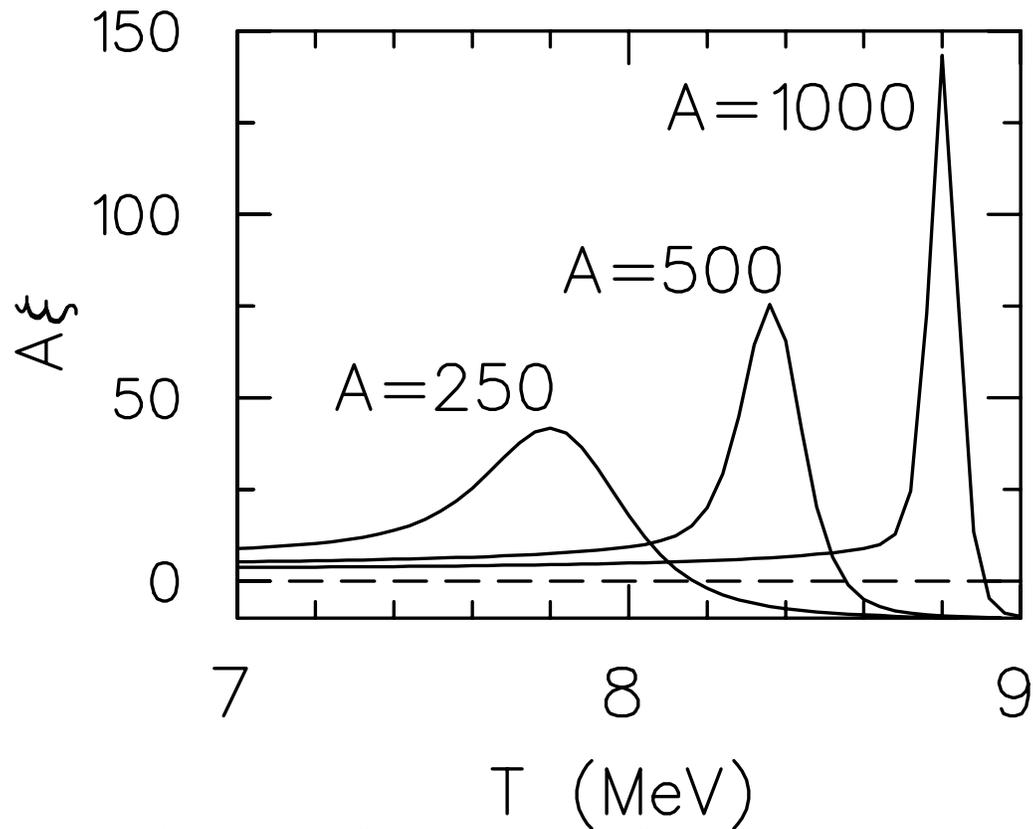}}
\caption{The evolution of the correlation coefficient is shown for a small
range of temperatures for increasing large system sizes. For large systems the
peak narrows into a singularity, a signal of discontinuous behavior. This is
remarkable given that it is the volume, not the pressure, that is held constant
while the temperature is changed. Coulomb has been neglected in this
calculation.
\label{transitionsharpness_fig}}
\end{figure}

The super-Poissonian (or nearly super-Poissonian when Coulomb is included)
behavior at the fragmentation threshold is especially interesting. Perhaps,
this may be understood by stating that the partition function is sampling two
competing configurations, one with one large fragment surrounded by gas and a
second one where numerous IMFs are present. At the fragmentation threshold,
where the system is undergoing a transition, both configurations occur leading
to a broadened multiplicity distribution. Figure \ref{transitionsharpness_fig}
displays $\xi$, again scaled by $A$, as a function of temperatures between 7
and 9 MeV, for 3 sizes, $A=250$, $A=500$ and $A=1000$. Coulomb is ignored in
these calculations. As the system size increases a singularity in $\xi$
develops at the boiling temperature. This is related to a discontinuity in the
energy vs. temperature at constant volume\cite{dasgupta}, which is
characteristic of this model but not characteristic of a typical liquid-gas
transition. In a liquid-gas transition $C_p$ becomes singular but not
$C_V$. Since the peak in $\xi$ vs. $T$ is linked to the discontinuity in the
energy density at the same temperature, the peak might disappear in a
microcanonical treatment.

\begin{figure}
\epsfxsize=0.75\textwidth \centerline{\epsfbox{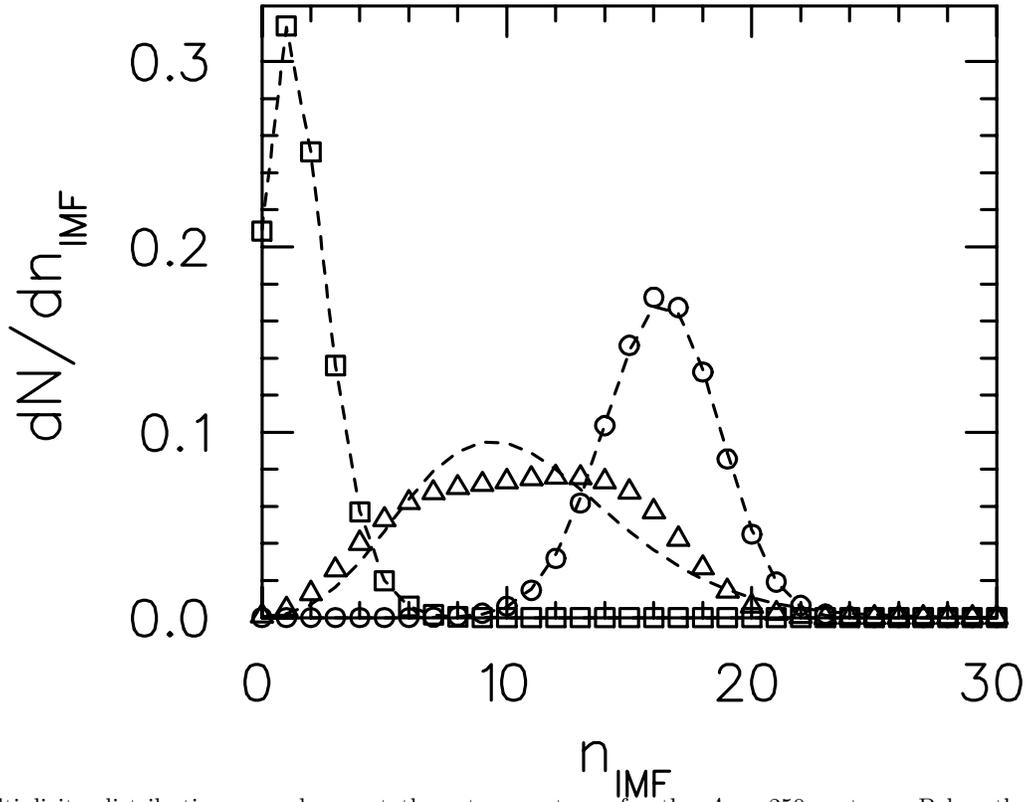}}
\caption{Multiplicity distributions are shown at three temperatures for the
$A=250$ system. Below the fragmentation threshold at a temperature of 7 MeV
(squares), the distribution is peaked at a low multiplicity and is well
described by a negative binomial distribution shown as a dashed line. At a high
temperature of 9 MeV (circles), the distribution is sub-Poissonian and well
described by the binomial distribution (dashed line). At 8 MeV (triangles), the
multiplicity distribution is strongly super-Poissonian and is not even well
described by a negative binomial.
\label{multdist_fig}}
\end{figure}

The multiplicity distributions for the $A=250$ system with Coulomb ignored are
shown in Figure \ref{multdist_fig} for three temperatures, 7, 8 and 9 MeV. They
are compared to negative binomial and binomial distributions respectively,
where the two parameters are chosen to fit the mean and variance of the
distributions. At 7 and 9 MeV, the two-parameter fits seem quite sufficient to
describe the multiplicity distributions, whereas at 8 MeV, where the
distribution is strongly super-Poissonian, the distribution's shape is poorly
described by a negative binomial fit. This emphasizes that two classes of
events, corresponding to the two phases, contribute at this temperature.

\section{Comparison with the percolation model}
\label{perc_sec}
Here, we present a brief comparison of the results of the canonical model
described in the previous sections with a percolation calculation\cite{gharib}.
For the percolation calculation, a cubic lattice of sites within a specified
radius of the center are attached by bonds. The bonds are then broken with a
probability $p$. When $p$ is in the neighborhood of $p_c=0.7512$, the system
undergoes a second order phase transition and the size of the average largest
cluster rapidly falls.

In the process of comparing percolation calculations to nuclear data, Li et
al.\cite{tongli} have developed a simple mapping of the bond breaking
probability with the temperature. By using this mapping we are able to plot the
results against temperature instead of $p$, and more readily compare to the
results of the statistical model presented earlier. In the calculation
presented here, the number sites in the spherical lattice was 203 and an IMF is
defined was a fragment with mass between 6 and 40.

\begin{figure}
\epsfxsize=0.75\textwidth \centerline{\epsfbox{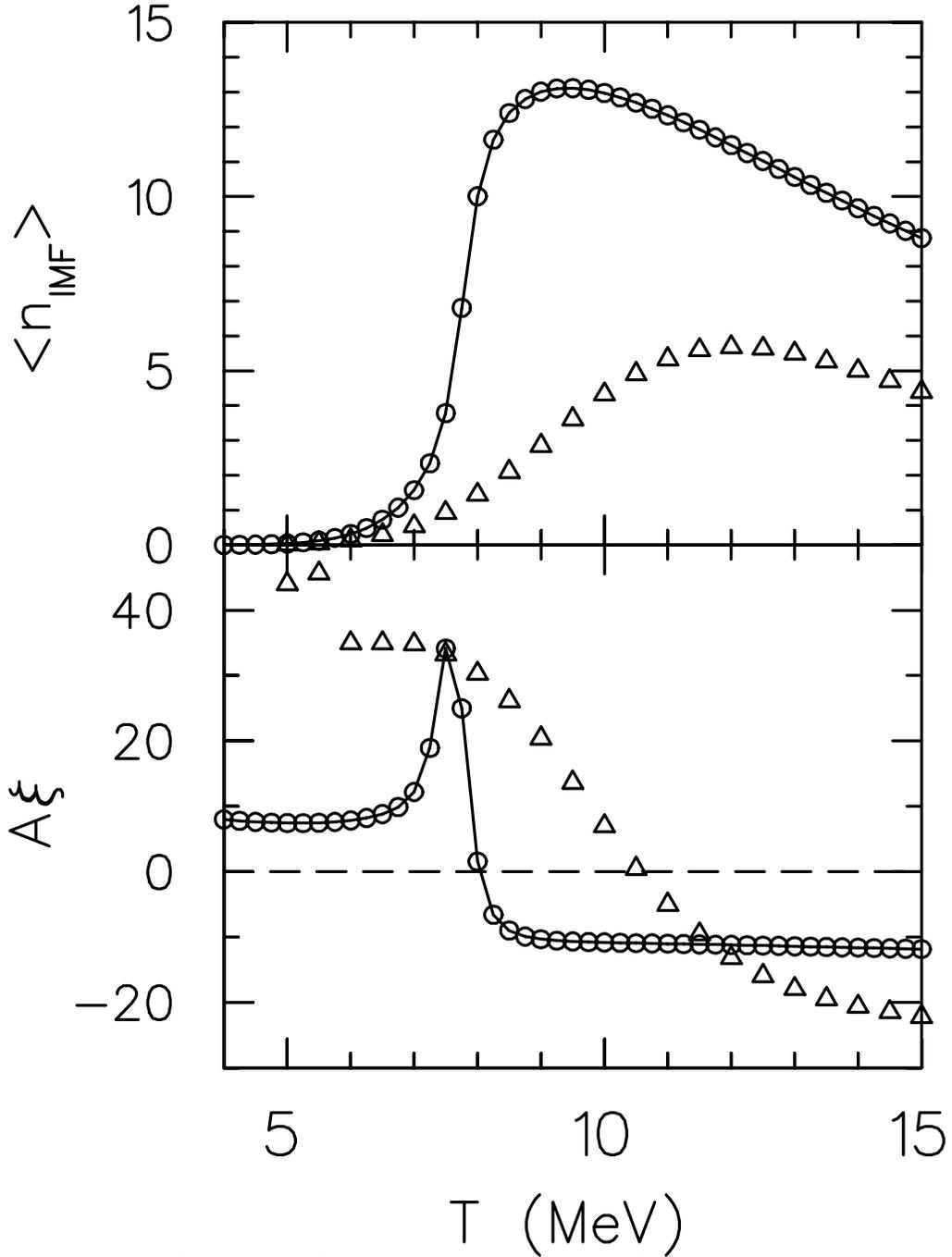}}
\caption{IMF multiplicity distributions are displayed in the upper panel, while
correlation coefficients are displayed in the lower panel. Percolation
calculations, based on a spherical lattice of 203 sites are represented by
triangles while the canonical ensemble, with $A=200$ and no Coulomb effects, is
represented by circles. The percolation model does not exhibit a first order
phase transition, hence it's correlation coefficient does not show a spike at
the critical temperature. Other notable differences are that the multiplicity
in the statistical calculation reaches higher values than in the percolation
model while the correlation coefficient in the percolation calculation reaches
larger values at low temperatures.}
\label{perc_fig}
\end{figure}

The mean IMF multiplicity and correlation coefficients are displayed in Figure
\ref{perc_fig}. They are contrasted to the results of the canonical-ensemble
calculations where $A$ was chosen as 200 and Coulomb was ignored. There are
three striking differences in the results. First, the percolation calculation
does not yield as many fragments as the statistical calculation, which produces
over twice as many fragments. Secondly, the percolation calculation does not
exhibit a spike in the correlation coefficient as plotted against the
temperature. This is expected as the percolation model does not contain a first
order phase transition. Finally, the correlation coefficient is much larger for
the case of the percolation calculation at low temperature. This is due to the
fragmentative nature of percolation, where the existence of a fragment opens
surface for the production of a second fragment\cite{gharib}.

\section{Microcanonical Calculations}
\label{microcanonical_sec}
A microcanonical approach, which considers configurations only at a specific
energy, would be more realistic since nuclear collisions do not take place with
contact to a heat bath. The importance of performing microcanonical
calculations is emphasized by the existence of the first order phase
transition, which collapses a wide range of energy densities to a narrow range
of temperatures. In this section we present expressions for calculating
fragmentation observables within a microcanonical context.

It is straightforward to obtain the needed microcanonical quantities from the
expressions for partition functions by Fourier transforming corresponding
canonical objects over a range of complex $\beta\equiv 1/T$. For instance, the
level density, $\rho(E)$, may be obtained from the partition functions through,
\begin{eqnarray}
\label{rhodef_eq}
\rho(E)&=&{\rm Tr}_\alpha\delta(E-E_\alpha)\\
\nonumber
&=&\frac{1}{2\pi}
\int d\beta \Omega_A(i\beta)e^{i\beta E}.
\end{eqnarray}

One may calculate recursion relations for the factorial moments
and for the multiplicity distributions at fixed energy. First we consider, 
$f_{b,A,m}(i\beta)$ and $p_{b,A,n}(i\beta)$ which are the canonical quantities
without the normalization brought on by dividing by $\Omega$. As described in
the appendix, they may be found via recursion relations.
\begin{eqnarray}
f_{b,A,m}(\beta)&\equiv&\Omega_A(\beta)F_{b,A,m}(\beta)\\
\nonumber
&=&\sum_{k\in b}\omega_k f_{b,A-a_k,m-1}(\beta)
\end{eqnarray}
\begin{eqnarray}
p_{b,A,n}(\beta)&\equiv&\Omega_A(\beta)p_{b,A,n}(\beta)\\ \nonumber
&=&\frac{1}{n}\sum_{k\in b}\omega_k p_{b,A-a_k,n-1}(\beta)
\end{eqnarray}

The factorial moments and multiplicity distributions at fixed energy $E$ may
then be expressed as,
\begin{eqnarray}
F_{b,A,m}(E)=\frac{\int d\beta f_{b,A,m}(i\beta)e^{i\beta E}}{2\pi\rho(E)}\\
\nonumber
P_{b,A,n}(E)=\frac{\int d\beta p_{b,A,n}(i\beta)e^{i\beta E}}{2\pi\rho(E)}
\end{eqnarray}

However, the integrations over $\beta$ needed to obtain the relevant
microcanonical quantities do make the calculations significantly more
numerically intensive. It is not clear to what degree the stationary-phase
approximation\cite{bohrmottleson} might allow one to avert the numerically
costly integration, especially given the first order phase transition which
causes discontinuities as a function of $\beta$.

If one is interested in the microcanonical distribution, for a range of
energies in the neighborhood of $E$, rather than for an exact value of $E$,
one may replace the delta function in Eq. (\ref{rhodef_eq}) by a Gaussian,
\begin{equation}
\delta(E-E_\alpha)\rightarrow 
\frac{1}{\sqrt{2\pi\eta^2}}\exp -\frac{(E-E_\alpha)^2}{2\eta^2},
\end{equation}
where $\eta$ defines the width of the neighborhood. One may incorporate the
broadening of the $\delta$ function by modifying the phase factors used in the
Fourier transforms,
\begin{equation}
e^{i\beta E}\rightarrow e^{i\beta E}\exp -\frac{1}{2}\eta^2\beta^2
\end{equation}
Numerical implementation of the Fourier transform simplifies for broader widths
$\eta$, as it effectively narrows the required integration range for $\beta$.

An alternative way to approach the constraint of energy conservation is to
discreetize the energy and measure it with integral values. One can then treat
energy in the same manner with which one would treat other conserved
charges. For instance, energy might by measured in steps of 0.1 MeV, with an
integer $Q$ measuring the energy. If $\omega_{i}$ counts the number of states
with energy $q_i$ and mass $a_i$, the number of states of the system of mass
$A$ with net energy $Q$ becomes,
\begin{equation}
N(A,Q)=\sum_{\langle \Sigma n_i a_i=A,\Sigma \langle n_iq_i=Q\rangle}
\Pi \frac{\omega_i^{n_i}}{n_i!} 
\end{equation}
One may then derive the recursion relation in a manner similar to the
derivation of the recursion relation for the partition function shown in
Section \ref{partfunc_recursion_subsec}. Since there are two conserved
``charges'', two recursion relations may be derived,
\begin{eqnarray}
N(A,Q)&=&\sum_i \frac{a_i}{A}\omega_iN(A-a_i,Q-q_i)\\
&=&\sum_i\frac{q_i}{Q}\omega_iN(A-a_i,Q-q_i)\\
\end{eqnarray}

In a manner similar to the derivations in Sections
\ref{factmom_recursion_subsec} and \ref{multdist_recursion_subsec} one may
derive recursion relations for the factorial moments and multiplicity
distributions,
\begin{eqnarray}
F_{b,A,Q,m}&=&\sum_{k\in b}
\omega_kF_{b,A-a_k,Z-Q_k,m-1}\frac{N(A-a_k,Q-q_k)}{N(A,Q)},\\
P_{b,A,Q,m}&=&\sum_{k\in b}
\omega_k\frac{1}{n}P_{b,A-A_k,Q-Q_k,n-1}\frac{N(A-a_k,Q-q_k)}{N(A,Q)}.
\end{eqnarray}

There are two practical differences between these expressions and the
corresponding canonical expressions. First, an extra index has been added that
denotes the energy of the system. Secondly, the index $k$ refers to a set of
states specified both by mass and charge. In practice this leads to a longer
calculation by a factor of the number of energy steps squared. Thus if one
wishes to perform a microcanonical calculation with an excitation energy of one
GeV, using energy steps of 1.0 MeV, the required computer time would be
expected to increase by $10^6$ as compared to a microcanonical calculation at a
single temperature.

\section{Conclusions}
\label{conclusions_sec}
The relations derived in this paper allow the exact calculation of all
fragmentation observables within the context of a canonical ensemble of
fragments. The relations allow the exact summation over all possible partitions
of $A$ nucleons within a fraction of a second of CPU time. The drawback of this
approach is that interactions between fragments may only be included in a
mean-field context, or through excluded volume. The advantage of this type of
approach goes beyond numerical convenience. Unlike many Monte Carlo approaches,
this technique has no arbitrary choices inherent to the algorithm. Thus, when
one has finished with a calculation, one can more clearly state what
expressions one has solved.

This approach can be made more realistic in three ways. First, the sum over
intrinsic states, which here was performed in a liquid-drop context, can be
replaced by a sum over true nuclear states without a significant cost in the
time of the calculation. Secondly, isospin conservation may be included. It is
straightforward to extend the recursion relations to two conserved
charges\cite{dasguptapratt} without significantly increasing the complexity of
the approach. Finally, a microcanonical ensemble would represent a
significantly more realistic description. Given that an entire region of energy
densities is described by a single temperature, the spikes observed in our
calculations of the correlation coefficient, and in previous
calculations\cite{dasguptapratt} of $C_V$, are expected to disappear in a
microcanonical approach. As discussed in the previous section, the methods
presented here might be extended to a microcanonical description, although
computational pitfalls may complicate numerical implementation of the
expressions derived here.

We finish by stating our belief that, although this type of approach is not yet
to the stage where it can seriously be compared to data, this represents a
revolution in the manner in which statistical physics can be applied to
predicting observables for finite systems. We foresee that this approach will
soon be developed to the stage where it can more conveniently, and more
transparently, provide insight into the interpretation of a variety of
experimental measurements.

\section{Appendix: Derivation of recursions relations used in canonical 
ensembles}
\subsection{Recursion relation for the partition function}
\label{partfunc_recursion_subsec}

The recursion relation described here was first presented by Lee and
Mekjian\cite{leemekjian}, and was first applied in a liquid-drop context by
Chase and Mekjian\cite{chase}.

The general relation for the partition function of non-interacting species is
\begin{equation}
\label{parteqdef_eq}
\Omega_A=\sum_{<\Sigma n_ka_k=A>}\prod_k \frac{\omega_k^{n_k}}{n_k!} ,
\end{equation}
where $\omega_k$ is the partition function for a single particle of the species
$k$ which has size $a_i$. For each term in the sum, one can factor a term
$\omega_k$ out of the partition function. By using the fact that $\Sigma
n_ka_k/A=1$, one may rewrite the partition function,
\begin{eqnarray}
\label{recursivealmost_eq}
\Omega_A&=&\sum_k\sum_{<\Sigma n_ia_i=A>}\frac{n_ka_k}{A}
\frac{\omega_k^{n_k}}{n_k!}\prod_{i\ne k} \frac{\omega_i^{n_i}}{n_i!}\\
&=&\sum_k \omega_k\frac{a_k}{A}\sum_{<\Sigma n_ia_i=A-a_k>}
\prod_{i} \frac{\omega_i^{n_i}}{n_i!}
\end{eqnarray}
From combining this expression with Eq. (\ref{parteqdef_eq}), one can extract
the recursion relation,
\begin{equation}
\Omega_A=\sum_k\omega_k \frac{a_k}{A}\Omega_{A-a_k}.
\end{equation}

\subsection{The recursion relation for factorial moments}
\label{factmom_recursion_subsec}
Factorial moments allow convenient calculation of the multiplicity distribution
as seen in the subsequent subsection. Given the partition function,
$\Omega_A$, the moments, $F_{k,A,m}$, are trivial to calculate for an
individual species, $k$.
\begin{eqnarray}
\label{factmomonespecies_eq}
F_{k,A,m}&\equiv&\langle n_k(n_k-1)(n_k-2)\cdots(n_k-m+1)\rangle\\
\nonumber
&=&\frac{1}{\Omega_A}\sum_{<\Sigma n_ia_i=A>}
n_k(n_k-1)\cdots(n_k-m+1)
\prod_{i} \frac{\omega_i^{n_i}}{n_i!}\\
\nonumber
&=&\omega_k^n\frac{\Omega_{A-ma_k}}{\Omega_A}.
\end{eqnarray}
However, they are more difficult to obtain when they are defined in terms of
$n_b$ comprised of several species with different masses,
\begin{equation}
n_b=\sum_{k\in b}n_k
\end{equation}
However in this case one may proceed with the help of a recursion relation for
the factorial moments. To derive the recursion relation, we consider the
function $f$.
\begin{eqnarray}
\label{littlefdef_eq}
f_{A,m}(b)&\equiv&\sum_{\langle \Sigma n_k=A\rangle}
\prod_k \frac{\omega_k^{n_k}}{n!}
n_b(n_b-1)\cdots (n_b-m+1)\\
F_{b,A,n}&=&\frac{f_{A,n}(b)}{\Omega_A}
\end{eqnarray}
For the first term $n_b=\sum_{k\in b}n_k$ in the sequence of $m$ terms 
$n_b(n_b-1)\cdots(n_b-m+1)$, each power of $n_k$ may be used to cancel $n_k$ in
the factorial. By then factoring $\omega_k$ outside the sum over
configurations, one may rewrite $f$ as
\begin{equation}
\label{factmom_recursion_b_eq}
f_{A,m}(b)=\sum_{k \in b}\omega_k
\sum_{\langle \Sigma n_{k^\prime}=A-a_{k^\prime}\rangle}
\prod_{k^\prime} \frac{\omega_{k^\prime} ^{n_{k^\prime}}}{n_{k^\prime}!}
n_b^\prime(n_b^\prime-1)(n_b^\prime-2)\cdots (n_b^\prime-m),
\end{equation}
where $n_b^\prime$ represents the number of $b$-type fragments in the set
$k^\prime$, which differs from the previous set by the reduction of one
fragment of type $k$.
From the definition of $f$, one may rewrite Eq. (\ref{factmom_recursion_b_eq}),
\begin{equation}
\label{factmom_recursion_c_eq}
f_{A,m}(b)=\sum_{k\in b}\omega_k f_{A-a_k,m-1},
\end{equation}
which leads to the recursive expression for $F$,
\begin{equation}
\label{factmom_recursion_a_eq}
F_{b,A,m}=\sum_{k\in b}\omega_k F_{b,A-a_k,m-1}\frac{\Omega_{A-a_k}}{\Omega_A},
\end{equation}
This recursion relation allows one to calculate factorial moments of increasing
order and for increasingly large nuclei given knowledge of the partition
function.

\subsection{Obtaining the multiplicity distribution from the factorial moments}
One can express the multiplicity distribution $P_{b,A,n}$ in terms of the
factorial moments. Here, $P_{b,A,n}$ is the probability of observing $n$
fragments of type $b$ in an event from a system of mass $A$.  The desired
expression, which we derive further below, has the simple form,
\begin{equation}
\label{factmomtomultdist_eq}
P_{b,A,n}=\sum_{m\ge n} F_{b,A,m}\frac{1}{(m-n)!n!}(-)^{m-n},
\end{equation}
where $F_{b,A,m}=\langle n_b(n_b-1)\cdots (n_b-m+1)\rangle$.  Only factorial
moments of order $n$ or greater contribute to $P_{b,A,n}$ since events with
multiplicity $n_b<m$ do not contribute to $F_{b,A,m}$.

To prove Eq. (\ref{factmomtomultdist_eq}) we rewrite the right-hand side of
Eq. (\ref{factmomtomultdist_eq}) using the definition of factorial moments,
\begin{eqnarray}
\sum_{m\ge n} F_{b,m} \frac{1}{(m-n)!n!}(-)^{m-n}&=&
\sum_{m\ge n} \sum_{\ell\ge m}
P_{b,A,\ell} \frac{\ell!}{(\ell-m)!}\frac{1}{(m-n)!n!}(-)^{m-n}\\
\nonumber
&=&\sum_{\ell\ge n}\sum_{m=0}^{\ell-n}
P_{b,A,\ell}\frac{\ell!}{(\ell-n-m)!}\frac{1}{m!n!}(-)^m,
\end{eqnarray}
where, in practice, the sums do not extend to $\infty$ due to the finite size
of the system.  The sum over $m$ can now be eliminated by using the identity,
\begin{equation}
\label{cuteidentity_eq}
\sum_{m=0}^k\frac{k!}{(k-m)!m!}(-)^m=(1-1)^k=\delta_{k,0},
\end{equation}
to obtain Eq. (\ref{factmomtomultdist_eq}). Although
Eq. (\ref{factmomtomultdist_eq}) is easy to implement numerically, it is
susceptible to problems with numerical accuracy due to the alternating
signs. Our experience is that such problems set in when the multiplicities
approach or exceed ten. However, a recursion relation for the multiplicity
distribution, which is derived in the next section, allows calculation of the
multiplicity distribution without first calculating the moments. Such an
expression does not have alternating signs and therefore is less susceptible to
numerical problems.

\subsection{The recursion relation for the multiplicity distribution}
\label{multdist_recursion_subsec}
In the previous sections of the appendix, relations have been derived that give
a recursion relation for the factorial moments, and also give the multiplicity
distribution in terms of the factorial moments. In this section we derive a
recursion relation for the multiplicity distribution, that will allow the
calculation of the multiplicity distribution without first calculating the
moments.

By inserting the recursion relation for factorial moments,
Eq. (\ref{factmom_recursion_a_eq}), into the formula for deriving the
multiplicity distribution in terms of factorial moments,
Eq. (\ref{factmomtomultdist_eq}), one obtains,
\begin{eqnarray}
P_{b,A,n}&=&\sum_{m\ge n}\frac{1}{(m-n)!n!}(-)^{m-n}\sum_{k\in b}\omega_k
F_{b,A-a_k,m-1}\frac{\Omega_{A-a_k}}{\Omega_A}\\
\nonumber
&=&\sum_{k\in b}\omega_k\frac{\Omega_{A-a_k}}{\Omega_A}
\sum_{m\ge 0}\frac{1}{m!n!}(-)^m F_{b,A-a_k,n+m-1}
\end{eqnarray}
By replacing $F_{b,A-a_k,n+m-1}$ in the above expression with it's definition
in terms of the multiplicity distribution,
\begin{eqnarray}
P_{b,A,n}&=&\sum_{k\in b}\omega_k\frac{\Omega_{A-a_k}}{\Omega_A}
\sum_{m\ge 0}\frac{1}{m!n!}(-)^m
\sum_{m^\prime\ge 0}P_{b,A-a_k,n+m+m^\prime-1}
\frac{(n+m+m^\prime-1)!}{m^\prime!)}\\
\nonumber
&=&\sum_{k\in b}\omega_k\frac{\Omega_{A-a_k}}{\Omega_A}
\sum_{M\ge 0}P_{b,A-a_k,n+M-1}\frac{(n+M-1)!}{n!}
\sum_{0\le m\le M} (-)^m \frac{1}{m!(M-m)!}\\
\nonumber
&=&\sum_{k\in b}\omega_k\frac{\Omega_{A-a_k}}{\Omega_A} 
P_{b,A-a_k,n-1}\frac{1}{n},
\end{eqnarray}
where the last step utilized the identity, Eq. (\ref{cuteidentity_eq}).

In practice, the multiplicity distributions are calculated for small $A$, then
for successively larger $A$ using the recursion relation above. However,
calculation of the $n=0$ term can not be determined from the recursion relation
and must be determined through the constraint, $\sum_n P_n=1$.

\acknowledgments{This work was supported by the National Science Foundation,
grant PHY-96-05207.}

\newpage

\end{document}